\documentclass[12pt]{article}
\usepackage{amssymb,amsmath}

\usepackage{graphicx}

\usepackage{amsfonts}

\makeatletter
\@addtoreset{equation}{section} 
\makeatother

\newcommand{\pd}[2]{\frac{\partial#1}{\partial#2}}
\newcommand{\beq}{\begin{equation}}
\newcommand{\eeq}{\end{equation}}
\newcommand{\barr}{\begin{eqnarray}}
\newcommand{\earr}{\end{eqnarray}}
\def\bsigma{{\mbox{\boldmath $ \sigma$}}}
\def\bpartial{{\mbox{\boldmath$\partial$}}}

\def\<#1,#2>{\left\langle#1,#2\right\rangle} 

\newcommand{\be}{\begin{equation}}
\newcommand{\ee}{\end{equation}}
\newcommand{\bea}{\begin{eqnarray}}
\newcommand{\eea}{\end{eqnarray}}

\begin{document}
\begin{titlepage}
\begin{flushright}
\baselineskip=12pt
DSF/18/2009\\
ICCUB-09-426
\end{flushright}

\begin{center}

\baselineskip=24pt

{\Large\bf Monopole-based quantization: a programme}

\baselineskip=14pt

\vspace{1cm}

{\bf Jos\'e F. Cari\~nena$^{a}$, J.M.\ Gracia-Bond\'{\i}a}$^{a}$,
{\bf Fedele Lizzi\,}$^{b,c}$, {\bf Giuseppe Marmo}$^{b}$ and {\bf
Patrizia Vitale}\,$^{b}$
\\[6mm] $^a$ {\it Departamento de F\'{\i}sica
Te\'{o}rica, Facultad de Ciencias,
\\
Universidad de Zaragoza, 50009 Zaragoza, Spain}
\\{\tt
jfc@unizar.es,jmgb@unizar.es}\\[6mm]
$^b$ {\it Dipartimento di Scienze Fisiche, Universit\`{a} di Napoli
{\sl
Federico II}
\\
and {\it INFN, Sezione di Napoli, Monte S.~Angelo}\\
Via Cintia, 80126 Napoli, Italy}\\
{\tt fedele.lizzi@na.infn.it, giuseppe.marmo@na.infn.it,
patrizia.vitale@na.infn.it}
\\[6mm]
$^c$ {\it High Energy Physics Group, Dept.\ Estructura i
Constituents de la Mat\`eria, \\Universitat de Barcelona, Diagonal
647, 08028 Barcelona, Catalonia, Spain \\and Institut de Ci\`encies
del Cosmos, UB, Barcelona}

\end{center}

\vskip 2 cm

\begin{abstract}
{We describe a programme to quantize a particle in the field of a
(three dimensional) magnetic monopole using a Weyl system. We
propose using the mapping of position and momenta as operators on a
quaternionic Hilbert module following the work of Emch and Jadczyk.}
\end{abstract}

\vskip 2 cm

\noindent{\sl Contribution to the volume: {\it Mathematical Physics
and Field
Theory, Julio Abad, In Memoriam}}\\
{\small M. Asorey, J.V.~Garc\'\i a Esteve, M.F. Ra\~nada and
J.~Sesma Editors, Prensas Universitaria de Zaragoza,~(2009)}

\end{titlepage}

\section{Introduction: several birds with a stone}
\label{sec:introibo}

\qquad Quantum kinematics in the field of a magnetic monopole allows
for angular momentum-isospin coupling, in particular spin-isospin
and orbit-isospin couplings.  However, to our knowledge the rich
conceptual and mathematical structures associated to angular
momentum-isospin coupling have gone unnoticed so far in deformation
quantization theory.

The aim of this contribution is to set out the basis for a rigorous
investigation of those structures in canonical quantization, up to
defining the pertinent monopole-based Weyl systems and star
products. We note that angular momentum-isospin coupling is a
feature of high physical interest; for a current example of
practical application, we refer to graphene edge
model~\cite{Tkachov}.  Rather than presenting novel results we will
outline, with the detail permitted by the total length limit, a
general framework in which it will be possible to use the monopole
for a variety of investigations both from the physical and
mathematical point of view.

A classical particle in the field of a magnetic monopole of unitary
charge is described, in proper units, by the Poisson structure:
\begin{eqnarray}
\{p_i,x^j\}&=&\delta_i^j\nonumber\\
\{x^i,x^j\}&=&0\nonumber\\
\{p_i,p_j\}&=&-\frac12 \epsilon_{ijk}\frac{x^k}{\|\mathbf{x}\|^2} \label{monopolePoisson}
\end{eqnarray}
This Poisson structure is position dependent and therefore its
quantization is not trivial, but extremely rich!  The right
mathematical framework for the quantization of this structure is the
formulation of quantum mechanics on a quaternionic Hilbert space
given by Emch and Jadcyzk over 10~years ago~\cite{EJ98}.  Since this
does not seem to be common knowledge, at some point we summarize the
findings of that paper, insofar they are useful for our purposes.

Prior to doing the above, we exhibit another problem of principle,
in some sense dual to quantization in the field of a monopole, whose
resolution hangs from the same mathematical thread.  Thus the
article is organized as follows.  An \textit{ab initio} calculation,
using the Kirillov coadjoint picture~\cite{Kirillov} allows to
regard the photon as a \textit{classical} elementary physical system
for the inhomogeneous Lorentz group~${\mathcal{P}}$.  On the arena
of phase space, this turns out to be formally dual (exchanging
position and momenta) to the orbit-isospin coupling system.
Section~3 deals with the structure of the gCCR (generalized
canonical commutation relations) on the quaternionic Hilbert space.
In Section~4, we show how the Emch--Jadcyzk (EJ) calculus provides
us the necessary tools for quantization of the above indicated
systems.  Quantization and dequantization proper are sketched in the
next section.  In Section~6 we give the conclusion and outlook for
construction of the star product describing the scalar
particle-monopole system.

\section{The orbit method for photons}
\label{sec:GanymedePoincare}

The unique Poincar\'e-invariant Stratonovich--Weyl (de)quantizer and
Moyal product on the phase space $T^*{\mathbb{R}}^3\times
{\mathbb{S}}^2\simeq{\mathbb{R}}^6\times{\mathbb{S}}^2$ for massive
relativistic particles with spin (degenerating to
$T^*{\mathbb{R}}^3$ for spinless particles) was constructed
in~\cite{Ganymede} with help of results in~\cite{Miranda}.  Its
practical interest is nil since (contrary to the Galilean case) it
breaks down as soon as one introduces interaction.  However, this
construction was an important matter of principle: the formalism
based on this Moyal product is equivalent to relativistic quantum
mechanics (and of course participates in its flaws).  In particular,
it gave geometric quantization on phase space and relativistic
Wigner functions, establishing the bridge between the Kirillov
coadjoint orbit picture and the Wigner theory of unitary irreps for
the Poincar\'e group~${\mathcal{P}}$.  For massless particles,
although we knew the arrival point as well (see~\cite{PrezOfCarolo}
for a modern, streamlined treatment), we were stumped.  The time has
come to revisit the question.

We describe coadjoint for the splitting
group~$\widetilde{\mathcal{P}}_0$ of the Poincar\'e group; this is
just the universal covering $T_4 \ltimes SL(2,{\mathbb{C}})$,
without nontrivial extensions~\cite{ConMarianico}.  The Lie algebra
of $\widetilde{\mathcal{P}}_0$ is generated by ten elements $H, P^i,
J^i, K^i$ (for $i = 1,2,3$) corresponding respectively to time
translations, space translations, rotations and pure boosts.  Write
elements of~$\widetilde{\mathcal{P}}_0$ in the standard form \beq g
= \exp(-a^0 H + {\mathbf{a}}\cdot{\mathbf{P}})\,
\exp(\zeta\,{\mathbf{n}}\cdot{\mathbf{K}})\,
\exp(\alpha\,{\mathbf{m}}\cdot{\mathbf{J}}),\nonumber \eeq where
$a=(a^0,{\mathbf{a}}) \in T_4$, ${\mathbf{n}}$ and ${\mathbf{m}}$
are unit 3-vectors, $\zeta \geq 0$ and $0 \leq \alpha \leq 2\pi$,
with the understanding that $\exp(2\pi{\mathbf{m}}\cdot{\mathbf{J}})
= -1_2$ in $SL(2,{\mathbb{C}})$ for all~${\mathbf{m}}$.  The
coadjoint action of~$\widetilde{\mathcal{P}}_0$
on~${\mathfrak{p}}^*$ can be derived from the well-known commutation
relations for the generators.  Let $h$ be the linear coordinate on
${\mathfrak{p}}^*$ associated to~$H\in\mathfrak{p}$, and similarly
let $p^i,j^i,k^i$ be the coordinates associated to $P^i,J^i,K^i$ ($i
= 1,2,3$).  The action in these coordinates is given in
Table~\ref{tbl:coad-action}.
\begin{table}[tbp]
\centering \caption{The coadjoint action ${\rm Coad\,}(\exp X) y$}
\vspace{6pt}
\begin{tabular}{l|llll}
\hline\hline
$X$ & $-a^0H$ & ${\mathbf{a}}\cdot{\mathbf{P}}$ &
$\alpha{\mathbf{m}}\cdot{\mathbf{J}}$ &
$\zeta{\mathbf{n}}\cdot{\mathbf{K}}$ \rule{0pt}{12pt} \\[\jot] \hline
$h$ & $h$ & $h$ & $h$ & $(\cosh\zeta) h - (\sinh\zeta)
{\mathbf{n}}\cdot{\mathbf{p}}$ \rule{0pt}{12pt} \\[\jot]
${\mathbf{p}}$ & ${\mathbf{p}}$ & ${\mathbf{p}}$ &
$R_{\alpha{\mathbf{m}}}\,{\mathbf{p}}$ & ${\mathbf{p}} - (\sinh\zeta)h
{\mathbf{n}} + (\cosh\zeta - 1)({\mathbf{n}}\cdot{\mathbf{p}})
{\mathbf{n}}$ \\[\jot] ${\mathbf{j}}$ & ${\mathbf{j}}$ & ${\mathbf{j}}
+ {\mathbf{a}}\times{\mathbf{p}}$ &
$R_{\alpha{\mathbf{m}}}\,{\mathbf{j}}$ & $(\cosh\zeta) {\mathbf{j}} +
(\sinh\zeta) {\mathbf{n}}\times{\mathbf{k}} - (\cosh\zeta -
1)({\mathbf{n}}\cdot{\mathbf{j}}) {\mathbf{n}}$ \\[\jot]
${\mathbf{k}}$ & ${\mathbf{k}} - a^0{\mathbf{p}}$ & ${\mathbf{k}} +
h{\mathbf{a}}$ & $R_{\alpha{\mathbf{m}}}\,{\mathbf{k}}$ &
$(\cosh\zeta) {\mathbf{k}} - (\sinh\zeta)
{\mathbf{n}}\times{\mathbf{j}} - (\cosh\zeta -
1)({\mathbf{n}}\cdot{\mathbf{k}}) {\mathbf{n}}$ \\[\jot] \hline\hline
\end{tabular}
\label{tbl:coad-action}
\end{table}

The orbits of this action arise as level sets of two {\it Casimir
functions} $C_1,C_2$ on~${\mathfrak{p}}^*$, which are easy to obtain
explicitly (or to guess from other treatments).  Let $p =
(h,{\mathbf{p}})$ be the {\it energy-momentum} 4-vector and $w =
(w^0,{\mathbf{w}})$ the {\it Pauli--Luba\'nski} 4-vector given by
\beq w^0 = {\mathbf{j}}\cdot{\mathbf{p}}, \qquad {\mathbf{w}} =
{\mathbf{p}}\times{\mathbf{k}} + h {\mathbf{j}}.\nonumber \eeq
{}From Table~\ref{tbl:coad-action} one verifies that $w^0$
transforms like $h$ and ${\mathbf{w}}$ like ${\mathbf{p}}$ under the
coadjoint action; in particular, under ${\rm
Coad\,}\bigl(\exp(\zeta{\mathbf{n}}\cdot{\mathbb{K}})\bigr)$:
\begin{eqnarray*}
w^0 &\mapsto &(\cosh\zeta) w^0 - (\sinh\zeta) {\mathbf{n}}\cdot{\mathbf{w}},
\\
{\mathbf{w}} &\mapsto& {\mathbf{w}} - (\sinh\zeta)w^0 {\mathbf{n}} +
(\cosh\zeta - 1)({\mathbf{n}}\cdot{\mathbf{w}}) {\mathbf{n}}.
\end{eqnarray*}
Thus the Casimir functions we seek are \beq C_1 := (pp) = - h^2 +
{\mathbf{p}}\cdot{\mathbf{p}}, \qquad C_2 := (ww) =
-({\mathbf{j}}\cdot{\mathbf{p}})^2 +
\|{\mathbf{p}}\times{\mathbf{k}} + h \,{\mathbf{j}}\|^2.\nonumber
\eeq Notice that $p$ and $w$ are orthogonal in the Minkowski sense:
$(pw) = 0$.  Let us focus on the shape of light-like coadjoint
orbits, for which $C_1=0$.  {For physical reasons} (no
`continuous-spin' representations) we take the momentous decision of
stipulating that $w$ is parallel to~$p$. Clearly ${\mathbf{p}} \in
{\mathbb{R}}^3 \setminus \{0\}$ (the origin is an orbit).  We can
postulate ${\mathbf{q}} := {\mathbf{k}}/h$, which is well defined,
and takes all values in ${\mathbb{R}}^3$, and everything is
determined:
\beq h
= \|{\mathbf{p}}\|, \qquad {\mathbf{p}} = {\mathbf{p}}, \qquad
{\mathbf{j}} = \lambda\,\frac{{\mathbf{p}}}{\|{\mathbf{p}}\|} +
{\mathbf{q}} \times {\mathbf{p}}, \qquad {\mathbf{k}} =
\|{\mathbf{p}}\|\,{\mathbf{q}},\nonumber
\eeq
where $\lambda{\mathbf{p}}/\|{\mathbf{p}}\|$ plays the role of the
spin, with the helicity $ \lambda =
{\mathbf{j}}\cdot{\mathbf{p}}/\|{\mathbf{p}}\|$ being the projection
of the total angular momentum ${\mathbf{j}}$ on the momentum.
Therefore the orbit is 6-dimensional, and isomorphic to
${\mathbb{R}}^3 \times ({\mathbb{R}}^3 \setminus \{0\})\simeq
{\mathbb{R}}^3 \times {\mathbb{R}}_+ \times {\mathbb{S}}^2$.  This
non-trivial topology has non-trivial consequences.

By the general theory, the Poisson bracket is given by
\beq
\{f, g\} = c_{ij}^{\!k} \pd f{x_i}\, \pd g{x_j}\, x_k.\nonumber
\eeq The
commutation relations for the generators yield:
\begin{equation}
\{p^i, q^j\} = \{p^i, h^{-1}k^j\} = h^{-1}\{p^i, k^j\} = -\delta_{ij}.
\label{eq:what-starts-well}
\end{equation}
On the other hand,
\begin{eqnarray}
\{q^i, q^j\} &=& \{h^{-1}k^i, h^{-1}k^j\} = h^{-2}\{k^i, k^j\} +
h^{-1}k^j\{k^i, h^{-1}\} + h^{-1}k^i\{h^{-1}, k_j\}
\nonumber \\
&=& h^{-2}(-\epsilon^{ij}_k j^k - q^jp^i + q^ip^j) =
-\lambda\epsilon^{ij}_k\frac{p^k}{\|{\mathbf{p}}\|^3};
\label{eq:sometimes-ends-badly}
\end{eqnarray}
which is dual to the Poisson structure (\ref{monopolePoisson}) upon
the exchange $p\leftrightarrow q$.

The coordinates in (\ref{eq:sometimes-ends-badly}) are \textit{not}
canonical coordinates (Darboux coordinates do not exist globally,
but $d^3{\mathbf{q}}\,d^3{\mathbf{p}}$ is a global Liouville
measure). All this agrees nicely with the analysis
in~\cite{balaboys}.  We can recover from Table~\ref{tbl:coad-action}
the expression of the coadjoint action
of~$\widetilde{\mathcal{P}}_0$ on the orbit in terms of the
coordinates $({\mathbf{p}},{\mathbf{q}})$.  There is no need to
rewrite the action on~${\mathbf{p}}$.  Also, we readily obtain:
\begin{eqnarray*}
\exp(-a^0 H) \,{\triangleright} {\mathbf{q}} &=& {\mathbf{q}} -
\frac{a_0{\mathbf{p}}}{\|{\mathbf{p}}\|}
\\
\exp({\mathbf{a}}\cdot{\mathbf{P}})\, {\triangleright} {\mathbf{q}} &=&
{\mathbf{q}} + {\mathbf{a}}
\\
\exp(\alpha{\mathbf{m}}\cdot{\mathbf{J}}) \,{\triangleright}
{\mathbf{q}} &=& R_{\alpha{\mathbf{m}}}{\mathbf{q}}.
\end{eqnarray*}
These formulae conform to our intuition as to how a relativistic
particle should behave.  They seem to indicate that, provided one
allows for non-commutativity, the `photon' (a massless relativistic
particle in general) is in some sense a localizable particle, since
the full Euclidean group is realized on a set of coordinate
variables. (\textit{Pace} the founding fathers: in the old
paper~\cite{cat-hater} Wightman writes that no such a realization
can exist; but he assumes commuting coordinates.)  The symplectic
form corresponding to~(\ref{eq:what-starts-well})
and~(\ref{eq:sometimes-ends-badly}) is given by: \beq \omega =
d{q}^i \wedge d{p}^i - \lambda \epsilon_{ijk}\frac{p^k\,dp^i\wedge
dp^j}{\|{\mathbf{p}}\|^3}.\nonumber \eeq This is exactly the one of
the magnetic monopole, with the roles of~${\mathbf{q}}$
and~${\mathbf{p}}$ interchanged: see further below. That is to say,
the `photon' and monopole phase spaces are dual systems.

The stability subgroup~$G_0$ giving rise to our coadjoint
orbit~$\mathcal{O}$ is a torus extension of the standard {\it little
group\/} for massless particles~$E_2$, so
$\widetilde{\mathcal{P}}_0/H\simeq {\mathbb{R}}^3 \times
{\mathbb{R}}_+ \times {\mathbb{S}}^2$.  However
$\widetilde{\mathcal{P}}_0/E_2\simeq{\mathbb{R}} ^3 \times
{\mathbb{R}}_+ \times {\mathbb{S}}^3$, and this~${\mathbb{S}}^3$
sits over~${\mathbb{S}}^2$ like in the Hopf~fibration.

\section{Quaternionic Weyl systems}
\label{sec:ultrameta}

Weyl systems on a complex Hilbert space~$\mathcal{H}$ are usually
presented starting with a (real) symplectic vector space, say
$(V,\omega)$, and a strongly continuous map
$V\to\mathfrak{U}(\mathcal{H})$ to the unitary group on it, required
to satisfy \beq W(v_1)W(v_2)W^\dag(v_1)W^\dag(v_2) =
e^{i\omega(v_1,v_2)}\mathbb{I}.  \eeq Strong continuity, by means of
the Stone-von Neumann theorem, implies that there exists a
selfadjoint operator $R(v)$ such that
\begin{equation}
W(v) = e^{iR(v)}  \label{expWeylnormal}.
\end{equation}
{}From the commutator relation we have \beq R\big([v_1,v_2]\big) -
[R(v_1),R(v_2)] = i\omega(v_1,v_2)\mathbb{I}.
\eeq Another theorem by von Neumann says that Weyl systems exist for
any finite dimensional symplectic vector space.  They can be defined
on a linear space of square integrable functions on any Lagrangian
subspace of~$\omega$ in~$V$.  If we denote by~$L$ such a Lagrangian
subspace, we may write \beq (V,\omega) \rightleftharpoons (L\oplus
L^*, \omega_0),\nonumber \eeq that is to say $(V,\omega)$ is
symplectically isomorphic with $(T^*L,d\theta_0\equiv\omega_0)$,
where $\theta_0$ is the Liouville 1-form on $T^*L$.  By denoting
$v\in V$ as $(y,\alpha)\in T^*L$, we have a Weyl system realized by
\beq
\begin{array}{rcl}
[W(0,\alpha)\psi](x) &=& e^{i\langle\alpha,x\rangle}\psi(x); \cr
[W(y,0)\psi](x) &=& \psi(x+y).
\end{array}
\eeq On defining either \beq W(y,\alpha) = W(y,0)W(0,\alpha) \qquad
{\rm or\ as}\qquad W(y,\alpha) = W(0,\alpha)W(y,0), \eeq we find an
{\it ordering\/} phase factor ambiguity.  At the infinitesimal level
we have a realization in terms of differential operators \beq i
R(y,0) = \frac{\partial}{\partial x}; \qquad i R(0,\alpha) = \hat x.
\eeq The symplectic structure evaluated on vectors $(y,0)$ and
$(0,\alpha)$ amounts to the commutator of the differential operators
${\partial}/{\partial x}$ and~$\hat x$.  In general, even though
differential operators are unbounded, one prefers to see the algebra
of operators acting on $\mathcal{H}$ realized as the algebra of
differential operators acting on functions defined on~$L$.  In this
framework it is more convenient to deal with square integrable
functions considered as sections of an associated $U(1)$-bundle $P$
over $L$.  It is possible to write a function $f$ on $L$ as a
function on $P$ by setting $\tilde f(x,t)= f(x) e^{it}$.  With this
choice functions on $L$ are associated with functions on $P$
satisfying the equation \beq -i \frac{\partial}{\partial t}\tilde f
= \tilde f; \eeq then our algebra of differential operators may be
realized in terms of vector fields, to wit ${\partial}/{\partial
x},-ix{\partial}/{\partial t}, -i\,{\partial}/{\partial t}$.  These
vector fields close the Lie algebra of the Heisenberg--Weyl group,
with$-i\,{\partial}/{\partial t}$ generating the linear space of
central elements.  While sections of a line bundle are appropriate
to describe spinless particles, to describe particles with an inner
structure it is necessary to consider sections of some Hermitian
complex vector bundle.  Usually we consider $f:L\rightarrow
{\mathbb{C}}^{2s+1}$ as functions associated with a trivialization
of the $U(2s+1)$ Hermitian bundle $ P$ over $L$. In this setting the
generators of our Weyl systems will be {\it matrix valued}
differential operators.

\smallskip

The approach to quaternionic Quantum Mechanics undertaken more than
forty years ago, may be considered from this perspective.
Let~${\mathbb{H}}$ denote the field of quaternionic numbers
\beq {\mathbb{H}}= \left\{q= \sum_{\mu = 0}^3 q^{\mu} e_{\mu} \mid
q^{\mu} \in {\mathbb{R}}\right\},\nonumber \eeq with their ordinary
multiplication and involution.  We will use the notations $1 = e_0$
and ${\mathbf{e}} = (e_1, e_2, e_3)$, so that
${\mathbf{x}}\cdot{\mathbf{e}} = \sum_{i=1}^3 x^i e_i$ for
${\mathbf{x}} \in {\mathbb{R}}^3$.  Note that $q^*q = \Vert
q\Vert_{\mathbb{H}}^2$ defines the quaternion norm, and that
$({\mathbf{x}}\cdot{\mathbf{e}})^*({\mathbf{x}}\cdot{\mathbf{e}}) =
\Vert{\mathbf{x}}\Vert^2$.  Then ${\mathcal{H}}\equiv
{\mathcal{L}}^2({\mathbb{R}}^3,d^3{\mathbf{x}}; {\mathbb{H}})$ is a
Hilbert space of quaternion-valued functions. We consider wave
functions realized on a Hilbert space (module) of quaternionic
valued functions.  By using the representation of quaternions by
means of $2\times 2$ skew-Hermitian matrices, we may write (when
convenient)
\beq e_0 = \sigma_0, \quad e_i = -i\sigma _i, \qquad{\rm and}\qquad
F(x) = f^0(x)e_0 + f^i (x) e_i.  \eeq The group $SU(2)$ acts on
these fibres by conjugation and the vector bundle may be considered
as an associated bundle with structure group $SU(2)$.  If we
identify $L\equiv{\mathbb{R}}^3$, we may repeat our construction and
lift vector fields from ${\mathbb{R}}^3$ to the total space of our
vector bundle.  To this aim we have to introduce a connection, that
is, a procedure to lift vector fields to horizontal vector fields.
We use the {\it gauge potential} \beq A = k \,
\frac{[{\mathbf{e}}\cdot{\mathbf{x}}, {\mathbf{e}}\cdot
d{\mathbf{x}}]}{\|\mathbf{x}\|^2}.  \eeq The origin of this choice
may be traced back to the Hopf fibration
$\pi:SU(2)\longrightarrow\mathbb{S}^2$: if we consider $s\in SU(2)$
we may define~\cite{OrangeOfYore}:
\beq
\pi(s) = s^{-1}\sigma_3 s = {\mathbf{x}}\cdot\bsigma \eeq and \beq
\begin{array}{rcl}
\bsigma
\cdot d{\mathbf{x}} &=& -s^{-1}ds s^{-1} \sigma_3 s + s^{-1}\sigma_3 ds
\nonumber \\
&=&-s^{-1}ds({\mathbf{x}}\cdot\bsigma) + ({\mathbf{x}}\cdot\bsigma)s^{-1}ds
\nonumber \\
&=& [s^{-1}ds, {\mathbf{x}}\cdot\bsigma].
\end{array}
\eeq
Moreover, since ${\mathbb{S}}^2\times
{\mathbb{R}}_+={\mathbb{R}}^3-\{0\}$, we may define a lifting which
would consider {\it wave functions\/} as fields transforming
covariantly under the rotation group, whose action in the inner
space is by means of~$SU(2)$.  Given any ${\mathbf{u}}\in
{\mathbb{S}}^2$, the translation
${\mathbf{u}}\cdot{\partial}/{\partial{\mathbf{x}}}$ is lifted to
\beq \nabla_{{\mathbf{u}}} =
e_0{\mathbf{u}}\cdot\frac{\partial}{\partial{\mathbf{x}}} +
\frac{1}{2} \frac{[ {\mathbf{e}}\cdot{\mathbf{x}},
{\mathbf{e}}\cdot{\mathbf{u}}]}{\|\mathbf{x}\|^2}, \eeq considered
as quaternionic-valued differential operator.  Clearly \beq
\nabla_{{\mathbf{u}}_1}\nabla_{{\mathbf{u}}_2} -
\nabla_{{\mathbf{u}}_2}\nabla_{{\mathbf{u}}_1} =
\Omega({\mathbf{u}}_1,{\mathbf{u}}_2) \eeq because \beq
\Big[{\mathbf{u}}_1 \cdot \frac{\partial}{\partial{\mathbf{x}}},
{\mathbf{u}}_2 \cdot \frac{\partial}{ \partial{\mathbf{x}}}\Big] =
0.
\eeq The curvature $\Omega$ is quaternion-valued and we may define the
three presymplectic forms \beq e_1\Omega= \omega_1,\qquad e_2\Omega=
\omega_2,\qquad e_3\Omega= \omega_3, \eeq giving us the gCCR.

\section{The Emch--Jadcyzk calculus}
\label{sec:CaballerosAndantes}

The EJ model is an appropriate quantum framework for orbit-isospin
coupling. In order to endow~${\mathcal{H}}$ with a complex linear
structure we introduce: for every ${\mathbf{x}}\neq 0$, let
$j({\mathbf{x}})$ be the imaginary unit~quaternion \beq
j({\mathbf{x}}) =
\frac{{\mathbf{e}}\cdot{\mathbf{x}}}{\Vert{\mathbf{x}}\Vert}.
\label{eq:j-denuestrosamores}
\eeq  Then the linear operator $J$
given by $(J\psi)({\mathbf{x}}) = j({\mathbf{x}})\psi({\mathbf{x}})$
satisfies the relations $J^*J = I = JJ^*$ and $J^* = - J$, that is,
it is unitary and skew--Hermitian; clearly, we also have $J^2 = -I$.
Remark that the choice (\ref{eq:j-denuestrosamores}) defines a $J$
invariant under rotations that commutes with the position operators;
this is an easy calculation using $L_i= \epsilon_{ijk}x_j\partial_k
- \frac 12 e_i$ for the generator of rotations.

\subsection{Noncommutative translations}
\label{sec:quod-erat-in-votis}

On $\mathcal{H}$ one usually defines the translation by $\mathbf{a}$
by the operator $V({\mathbf{a}})$ such that \beq
[V({\mathbf{a}})\psi]({\mathbf{x}}) = \psi({\mathbf{x}} -
{\mathbf{a}}),\nonumber \eeq but taking into account the character
of rays rather than vectors of states one can also admit a {\it
phase\/} factor.  Here such translation is realized by the operator
$U({\mathbf{a}})$ defined by: \beq
[U({\mathbf{a}})\psi]({\mathbf{x}}) = w({\mathbf{a}};{\mathbf{x}} -
{\mathbf{a}})\psi({\mathbf{x}} - {\mathbf{a}}), \label{eq:U}
\eeq with ${\mathbf{a}}\in{\mathbb{R}}^3$.  Here, for every
${\mathbf{a}}$, $w({\mathbf{a}};{\mathbf{x}})$ is the quaternion
\beq w({\mathbf{a}};{\mathbf{x}}) = \cos(\alpha/2) + j({\mathbf{x}}
\times {\mathbf{a}})\sin(\alpha/2) = \exp\left[j({\mathbf{x}} \times
{\mathbf{a}})\alpha/2\right], \label{eq:w}
\eeq with $\alpha$ being the angle between ${\mathbf{x}}$ and $
{\mathbf{x}} + {\mathbf{a}}$.  If we use $w$ to define the linear
operator $W({\mathbf{a}})$ by \beq
[W({\mathbf{a}})\psi]({\mathbf{x}}) =
w({\mathbf{a}};{\mathbf{x}})\psi({\mathbf{x}}), \quad \rm{a.e.} \eeq
and then \beq U({\mathbf{a}}) = V({\mathbf{a}})W({\mathbf{a}}).
\eeq

Some properties of $w$ which will be useful below are:
\begin{itemize}
\item $w(0;{\mathbf{x}})= 1$.
\item $w({\mathbf{a}};{\mathbf{x}})
    w^*({\mathbf{a}};{\mathbf{x}})=1$.
\item $w({\mathbf{a}};{\mathbf{x}} - {\mathbf{a}})=
    w^*(-{\mathbf{a}}; {\mathbf{x}})$.
\item $w(t{\mathbf{a}}; {\mathbf{x}} +
    s{\mathbf{a}})w(s{\mathbf{a}};{\mathbf{x}}) = w((s +
    t){\mathbf{a}};{\mathbf{x}})$.
\end{itemize}

Let now ${\mathbf{u}}\in{\mathbb{S}}^2$.  We can define generators
for the continuous unitary representation $U(s{\mathbf{u}})$ by \beq
\nabla_{{\mathbf{u}}} = \lim_{t\downarrow 0} \Biggl[\frac{d}{dt}
U(t{\mathbf{u}})\psi({\mathbf{x}}) \Biggr] = \lim_{t\downarrow 0}
\Biggl[\frac{d}{dt} w(t{\mathbf{u}};{\mathbf{x}} -
t{\mathbf{u}})\psi({\mathbf{x}} - t{\mathbf{u}})\Biggr].
\label{eq:pgen}
\eeq One obtains \beq \nabla_{{\mathbf{u}}} = \biggl({\mathbf{u}}
\cdot { \bpartial} + \frac 12 {\mathbf{e}}\cdot\frac{{\mathbf{u}}
\times
{\mathbf{x}}}{\vert\vert{\mathbf{x}}\vert\vert^2}\biggr),\nonumber
\eeq whereupon we recognize the sum of the (non-commuting)
infinitesimal generators of $V$ and $W$, respectively.  Thus, with
the obvious meaning for the $X_i$, we readily compute the following
commutation relations \beq
\begin{array}{rcl}
 [\nabla_i, X_j] &=& \delta_{ij},\nonumber
\\
{[} \nabla_i, \nabla_j{]} &=& -\frac 12 J \, \epsilon_{ijk}
{\displaystyle\frac{x^k}{\Vert {\mathbf{x}}\Vert^3}},
 \\
{[}X_i, X_j{]} &=& 0,
\nonumber
\end{array}
\eeq
that should be compared with~\eqref{monopolePoisson}
and~\eqref{eq:sometimes-ends-badly}.

\smallskip

The key result of the EJ calculus is that the so defined operators
$U({\mathbf{a}})$ actually define a locally operating projective
representation of the translation group \cite{LOR}, i.e.
\[
U({\mathbf{a}})U({\mathbf{b}}) = U({\mathbf{a}} +
{\mathbf{b}})M({\mathbf{a}},{\mathbf{b}}).
\]
Here $M({\mathbf{a}},{\mathbf{b}})$ is  a phase factor
multiplication
\[[M({\mathbf{a}},{\mathbf{b}})\psi]({\mathbf{x}})=m({\mathbf{a}},{\mathbf{b}},
{\mathbf{x}})\,\psi({\mathbf{x}})
\]
with $m({\mathbf{a}},{\mathbf{b}}, {\mathbf{x}})$ being given by
\[
m({\mathbf{a}},{\mathbf{b}},{\mathbf{x}}) =
\exp\big(JS({\mathbf{a}},{\mathbf{b}},{\mathbf{x}})\big),
\]
where $S$ denotes the (product of the monopole strength) by the area
of the triangle spanned by ${\mathbf{x}}, {\mathbf{x}} +
{\mathbf{a}}, {\mathbf{x}} + {\mathbf{a}} + {\mathbf{b}}$.  This
guarantees associativity: $
U({\mathbf{a}})\big[U({\mathbf{b}})U({\mathbf{c}})\big] =
[U({\mathbf{a}})U({\mathbf{b}})]U({\mathbf{c}})$ (see next section).

Except for the presence of the quaternionic complex structure~$J$,
this is similar to the Moyal product, which bodes well for the
quantization/dequantization procedure.

\section{Exponential representation of the Weyl system}
In the usual case Weyl systems are represented as exponential as in
(\ref{expWeylnormal}). It is useful to express also the quaternionic
Weyl system as an exponential. A first problem is that in the
quaternionic context there is no single imaginary unit. Nevertheless
EJ have provided the solution of the problem in the function
$j({\mathbf{x}})$ defined in (\ref{eq:j-denuestrosamores}), and its
operatorial counterpart $J$. We can therefore define the operator
\begin{equation}
P_i=J\nabla_i=-J\left( \partial_i + \frac{1}{2} \frac{\epsilon_{ijk} x_j e_k}{\| \bold x \|^2}\right)
\end{equation}
it is possible to prove that $J$ commutes with $\nabla_i$ and
therefore $[P_i,P_j]=(1/2) \epsilon_{ijk} (x^k/\|{\mathbf{x}}\|^3)$.
Therefore $P_i$ is the generator of translations in the quaternionic
Hilbert space. Notice that the two summands  in $P_i$ do not
commute.

Finally, let us consider the product of two finite translations
\begin{equation}
(U({\bold a}) U({\bold b})\psi)({\bold x})= (U({\bold a}) \psi')({\bold x})=
w({\bold a}; {\bold x}-{\bold a}) \psi'({\bold x}-{\bold a})=w({\bold a}; {\bold
x}-{\bold a})w({\bold b};{\bold x}-{\bold a}-{\bold b}) \psi({\bold x}-{\bold
a}-{\bold b}).\nonumber
\end{equation}
On the other hand,
\begin{equation}
(U({\bold a} + {\bold b}) M({\bold a}, {\bold b}) \psi)({\bold x})=w({\bold a} +
{\bold b}; {\bold x} - {\bold a} -{\bold b})(M({\bold a}, {\bold b}) \psi)({\bold
x}-{\bold a}- {\bold b}),
\end{equation}
with $M({\bold a},{\bold b})$ defined as
\begin{equation}
(M({\bold a}, {\bold b}) \psi)({\bold x}) = m({\bold a},{\bold b};{\bold x})
\psi ({\bold x})= w^*({\bold a} + {\bold b}, {\bold x} )w({\bold a}; {\bold x}
+{\bold b}) w( {\bold b}; {\bold x} )\,\psi({\bold x}).
\end{equation}
Since $w({\bold a};{\bold x})=1$ and $w(0;{\bold x-a})=w^*({\bold
a};{\bold x})$ we have that
$m({\mathbf{a}},{\mathbf{b}};{\mathbf{x}})$ satisfies
\begin{equation}
 m({\mathbf{a}},-{\mathbf{a}};{\mathbf{x}})=1 \label{prom}.
\end{equation}
We obtain then
\begin{equation}
U({\bold a}) U({\bold b})=U({\bold a} + {\bold b}) M({\bold a}, {\bold b}).
\label{transprod}
\end{equation}

We now construct a Weyl system  from $\mathbf{P}$ and $\mathbf{X}$.
Consider the operator
\begin{equation}
T(\alpha)= e^{J[ {\mathbf{a}}\cdot \mathbf{P}+ {\mathbf{a}'}\cdot\mathbf{X}]}=
e^{J {\mathbf{a}}\cdot {\mathbf{P}}} e^{J
  {\mathbf{a}}'\cdot\mathbf{X}} e^{\frac{1}{2}a^i a^{\prime j}
[P^i, X^j]}=e^{J {\mathbf{a}}\cdot{\mathbf{P}}} e^{J {\mathbf{a}}'\cdot{\mathbf{X}}}e^{-\frac{1}{2}J {\mathbf{a}}\cdot {\mathbf{a}}'} =e^{J {\mathbf{a}}'\cdot{\mathbf{X}}} e^{J {\mathbf{a}}\cdot {\mathbf{P}}} e^{\frac{1}{2}J {\mathbf{a}} \cdot {\mathbf{a}}'}\label{pqexp}
\end{equation}
with $\alpha=({\mathbf{a}},{\mathbf{a}}')$. Remember that $ \exp ( J
{\mathbf{a}}\cdot {\mathbf{P}}) \equiv U({\mathbf{a}})$. We have
then
\begin{equation}
(T(\alpha) \psi)({\mathbf{x}})=(e^{J {\mathbf{a}}\cdot {\mathbf{P}}} e^{J {\mathbf{a}}'\cdot{\mathbf{X}}}e^{-\frac{1}{2}J {\mathbf{a}} \cdot {\mathbf{a}}'}\psi)({\mathbf{x}})= w({\mathbf{a}};\mathbf{x-a})e^{j(\mathbf{x-a}) {\mathbf{a}}'\cdot(\mathbf{x-a})}e^{-\frac{1}{2}j(\mathbf{x-a})  {\mathbf{a}} \cdot {\mathbf{a}}'}\psi(\mathbf{x-a}),
\end{equation}
but also
\begin{equation}
(T(\alpha) \psi)({\mathbf{x}})=(e^{J {\mathbf{a}}'\cdot{\mathbf{X}}} e^{J {\mathbf{a}}\cdot {\mathbf{P}}} e^{\frac{1}{2}J {\mathbf{a}} \cdot {\mathbf{a}}'}\psi)({\mathbf{x}})=e^{j({\mathbf{x}}) \mathbf{a'x}}w({\mathbf{a}}; \mathbf{x-a})e^{\frac{1}{2}j(\mathbf{x-a}) {\mathbf{a}} \cdot {\mathbf{a}}'}\psi(\mathbf{x-a})\,.
\end{equation}
We compute
\begin{equation}
T(\alpha) T(\beta)=e^{J[{\mathbf{a}}\cdot {\mathbf{P}}+ {\mathbf{a}}'\cdot{\mathbf{X}}]}e^{J[ {\mathbf{b}} \cdot{\mathbf{P}}+ {\mathbf{b}}'\cdot{\mathbf{X}}]}=e^{J {\mathbf{a}}\cdot {\mathbf{P}}} e^{J {\mathbf{a}}'\cdot{\mathbf{X}}} e^{J {\mathbf{b}}\cdot
{\mathbf{P}}} e^{J {\mathbf{b}}'\cdot{\mathbf{X}}}e^{-\frac{1}{2}J({\mathbf{a}} \cdot {\mathbf{a}}'+{\mathbf{b}} \cdot {\mathbf{b}}')}\,.
\end{equation}
On using (\ref{transprod}) and (\ref{pqexp}) we arrive at the Weyl
system
\begin{equation}
T(\alpha) T(\beta)=T(\alpha+\beta)M({\mathbf{a}},{\mathbf{b}})
\exp\Bigl(\frac{1}{2}J ({\mathbf{a}}\cdot {\mathbf{b}}'-{\mathbf{b}}\cdot
{\mathbf{a}}')\Bigr)\,. \label{weylsys}
\end{equation}

We see that this Weyl system provides as usual, but there are two
phases. The first is the antisymmetric product of the two vectors in
$\mathbb{R}^6$. This would be present also in the absence of the
monopole, it gives the noncommutativity of position and momenta,
however with the `imaginary' unit given by the quaternionic radial
functions $j({\mathbf{x}})$. The factor $M$ instead is the one which
contains the information on the noncommutativity of the
translations. Both phases are of course crucial for the description
of the particle in the field of a monopole.

\section{Outlook}
Quaternions are well suited to describe the quantization of
classical particles in the presence of the field of a magnetic
monopole.  We have laid the basis for this quantization.  It is in
principle possible (and will be presented elsewhere) to construct a
full deformation quantization.  One can built a full Weyl map which
associates functions on phase space with operators on the
quaternionic Hilbert space.  The quantized functions belong to a
subalgebra, in such a way that the map is (at least formally)
invertible, and therefore provides a star product.  Possible
colour-breaking phenomena
---see~\cite{SixBraveMen,{MeatMIT}}--- are to be fit in the formalism; indeed,
there are plenty of questions here we should have answered long ago.
But ``the gaps in the knowledge of the wise has been filled even so
slowly''~\cite{NineForMortalMen}.

\noindent{\bf Note Added} After this work had appeared we
constructed the star product quantizing the motion of particle in a
monopole field in~\cite{Ravel}.

\noindent{\bf Acknowledgments:} FL~would like to thank the
Department of Estructura i Constituents de la materia, and the
Institut de Ci\`encies del Cosmos, Universitat de Barcelona for
hospitality. His work has been supported in part by CUR Generalitat
de Catalunya under project 2009SGR502.



\end{document}